\documentclass[aps,longbibliography,twocolumn,showpacs,amsmath,amssymb,floatfix,superscriptaddress]{revtex4-1}
\usepackage{color}
\usepackage{graphicx}

\begin{document}

\bibliographystyle{utcaps}
\title{Entanglement Spectra of Interacting Fermions in Quantum Monte Carlo Simulations}
\author{\firstname{Fakher F.} \surname{Assaad}}
\affiliation{Institut f\"ur Theoretische Physik und Astrophysik, Universit\"at W\"urzburg, Am Hubland, D-97074 W\"urzburg, Germany}
\author{\firstname{Thomas C.} \surname{Lang}}
\affiliation{Department of Physics, Boston University, Boston, MA 02215, USA}
\author{\firstname{Francesco} \surname{Parisen Toldin}}
\affiliation{Institut f\"ur Theoretische Physik und Astrophysik, Universit\"at W\"urzburg, Am Hubland, D-97074 W\"urzburg, Germany}
\date{\today}

\begin{abstract}
In a recent article T. Grover [Phys. Rev. Lett. {\bf 111}, 130402 (2013)] introduced a simple method to compute Renyi entanglement entropies in the realm of the auxiliary field quantum Monte Carlo algorithm. Here, we develop this approach  further and provide a stabilization scheme to compute higher order Renyi entropies and an extension to access the entanglement spectrum. The method is tested on systems of correlated topological insulators.
\end{abstract}

% http://www.aip.org/pacs/pacs2010/individuals/pacs2010_regular_edition/alpha_index.html
%02.70.Ss Quantum Monte Carlo methods
%03.67.-a Quantum information 
%71.10.-w Theories and models of many-electron systems
%73.43.-f Quantum Hall effects

\maketitle

%============================================================================================================
\section{Introduction}
%============================================================================================================

The concept of entanglement spectra and entropies is emerging as a powerful tool to access universal quantities such as the central charge of an associated conformal field theory and characterize ground states and their degeneracies beyond the Landau paradigm. Typically, the ground state of a local Hamiltonian is entangled over short distances. This short range entanglement leads to an area law (or more precisely, a perimeter law) of the Renyi and von Neumann entanglement entropies \cite{Eisert10}. Corrections to the area law have implications for the nature of the ground state. For instance, topologically ordered states in two dimensions with anyonic elementary excitations \cite{Kitaev06} are characterized by a sub-leading correction to the area law \cite{Levin06,Kitaev06_1}. This signature of topological order has become a numerical tool that has been implemented within the frameworks of variational quantum Monte Carlo (QMC) \cite{Grover11}, stochastic series expansions \cite{Hastings10,Isakov11}, and density matrix renormalization group (DMRG) \cite{Yan11,Jiang12,Depenbrock12}. Other corrections to the area law include sub-leading logarithmically diverging corrections which signalize the presence of Goldstone modes of spontaneously broken continuous symmetries \cite{Metlitski11}. Entanglement entropies have also been introduced in the realm of the valence bond quantum Monte Carlo simulations for spin systems \cite{Sandvik05,Beach06,Alet07}. The Renyi entropy itself does not necessarily provide an efficient tool to characterize states of matter. In fact, and as pointed out recently in \cite{Budich13}, there is no practical way of distinguishing a Chern insulator from a trivial band insulator by computing the entanglement entropy. 

More information on the nature of the ground state is contained in the entanglement spectrum. The entanglement spectrum picks up the edge physics of topologically ordered fractional quantum Hall states \cite{Haldane08,Lauchli10,Thomale10,Qi12} as well as that of topological insulators \cite{Fidkowski10,Turner10}. In magnetically ordered states with spontaneously broken SU(2) spin symmetry, Anderson's tower of states can be detected in the entanglement spectrum \cite{Kolley13}. In a recent work Chandran {\it et al.} caution against the unconditional assumption that the {\it low energy } sector of the entanglement spectrum
contains universal properties of the state and of quantum phase transition \cite{Chandran13}.

The aim of this article is to extend a recent proposal to compute Renyi entropies in the realm of the auxiliary field QMC algorithm \cite{Grover13}, in order to access the entanglement spectrum. In the next section, we will set the stage and briefly review the zero temperature formulation of the auxiliary field QMC method as well as the calculation of Renyi entropies. Emphasis will be placed on stability issues we encountered when computing higher order Renyi entropies on {\it large} subsystems. Our approach to compute the entanglement spectrum is based on spectroscopy along the replica time axis. This notion will be introduced in Section \ref{ES.sec}. We test the method by computing the single-particle entanglement spectral function for the Kane-Mele Hubbard model supplemented by a single-particle bond dimerization term \cite{Lang13_1}. This dimerization triggers a topological phase transition to a topologically trivial state, which manifests plainly in the entanglement spectrum. In section \ref{Conclusions.sec} we conclude and discuss the shortcomings and strong points of our approach.

%============================================================================================================
\section{Renyi entropies from the zero temperature auxiliary field QMC method}
%============================================================================================================
%------------------------------------------------------------------------------------------------------------
\subsection{The Projective auxiliary field QMC algorithm}
%------------------------------------------------------------------------------------------------------------

\begin{figure}
   \includegraphics[width=0.8\linewidth]{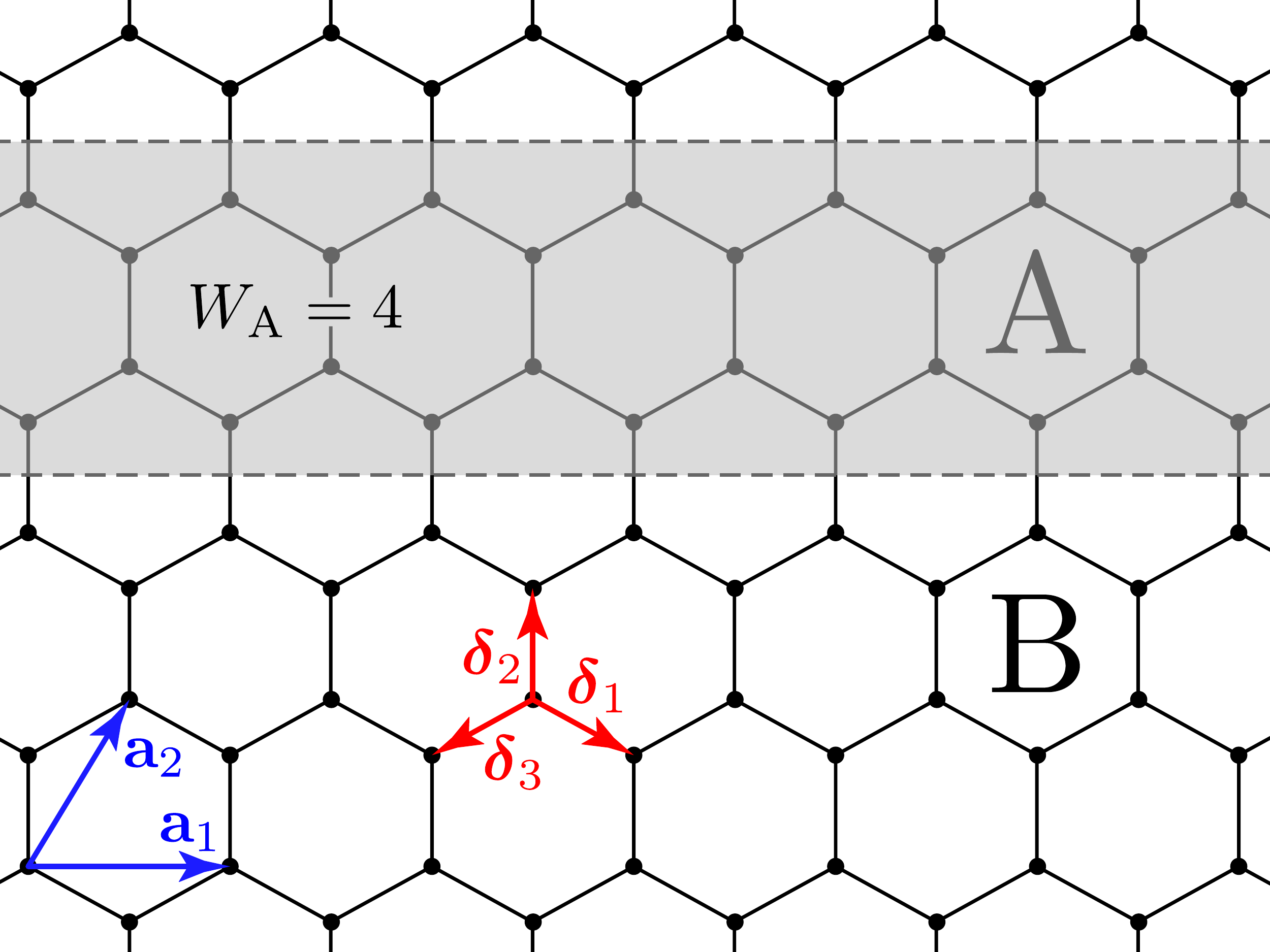} 
   \caption{Honeycomb lattice and the regions A and B. In the figure region A has a width of $W_\text{A} = 4$.
   \label{Latt.fig}}
\end{figure}

Here we provide a short description of the zero temperature, or projective, auxiliary field QMC (PQMC) algorithm for the Kane-Mele Hubbard model:
\begin{equation}
\label{Hamiltonian.eq}
	\hat{H} = \hat{{\mathbf{c}}}^{\dagger} \mathbf{T} \hat{\mathbf{c}} + \frac{U}{2}\sum_{\pmb{i}} \left(\hat{n}_{\pmb{i}} - 1\right)^2 \equiv \hat{H}_T + \hat{H}_U\;.
\end{equation}
$\hat{\mathbf{c}}^{\dagger} $ is a vector of fermionic creation operators with entries $\hat{c}_{\pmb{i}\sigma}^{\dagger}$. The operator $\hat{c}_{\pmb{i}\sigma}^{\dagger}$ creates an electron at lattice site $\pmb{i}$ with a $z$-component of spin $\sigma$, and $\hat{n}_{\pmb{i}} = \sum_{\sigma} \hat{c}_{\pmb{i}\sigma}^{\dagger}\hat{c}_{\pmb{i}\sigma}^{\phantom{\dagger}}$ is the charge density at site $\pmb{i}$. The kinetic energy is defined by the $N\times N$ Hermitian matrix $\mathbf{T}$, and $N$ corresponds to the number of single-particle states, which we will henceforth label with super-index $x \equiv\left( {\pmb i}, \sigma \right)$. In this article, $\hat{H}_T$ will be taken to be the dimerized Kane-Mele model. Using the spinor notation ${\hat{{\mathbf{c}}}^{\dagger}_{i} = \big(\hat{c}^{\dagger}_{\pmb{i}\uparrow}, \hat{c}^{\dagger}_{{\pmb i}\downarrow}\big)}$,
\begin{align}
\label{eq:KM}
  \hat{H}_{T} =
   \sum_{ {\pmb i},{\pmb j} } \hat{c}^{\dagger}_{\pmb{i}} \left[ t_{ \pmb{i}\pmb{j} } + {\rm i}\, \boldsymbol{\lambda}_{{\pmb i}{\pmb j }} \cdot \boldsymbol{\sigma} \right] \hat{c}^{\phantom{\dag}}_{\pmb j}\;.
 \end{align}
The hopping matrix takes non-vanishing values between nearest neighbors of the honeycomb lattice, ${{\pmb i } - {\pmb j} = \pm{\pmb \delta}_1, \pm{\pmb \delta}_2, \pm{\pmb \delta}_3}$ (see Fig.~\ref{Latt.fig}), and we have implemented the following dimerization: 
\begin{equation}
	t_{ \pmb{i}\pmb{j} } = 
   \left\{
      \begin{array}{cl}
         -t  & \text{if } \pmb{i} - \pmb{j} = \pm {\pmb \delta}_2, \pm{\pmb\delta}_3 \\
         -t' & \text{if } \pmb{i} - \pmb{j} = \pm {\pmb \delta}_1 \\
         0   & \text{otherwise}
      \end{array}
   \right..
\end{equation}
The intrinsic spin-orbit term is given by
\begin{equation}
	{\pmb \lambda}_{ \pmb{i}\pmb{j} } = \lambda
   \left\{
      \begin{array}{cl}
         \frac{ (\pmb{ i} - \pmb{r}) \times (\pmb{ r} - \pmb{j}) }{ \left|(\pmb{ i} - \pmb{r}) \times (\pmb{ r} - \pmb{j}) \right| } & \text{if } \pmb{i},\pmb{j} \text{ are n.n.n.} \\
         0 & \text{otherwise} 
      \end{array}\;,
   \right.
\end{equation} 
where ${\pmb r} $ is the intermediate site involved in the next nearest neighbor (n.n.n.) hopping process from site $ \pmb{i} $ to $\pmb{j}$. 

The PQMC algorithm is based on filtering out the ground state from a Slater determinant trial wave function: 
\begin{equation}
    | \Psi_\text{T}\rangle = \prod_{n=1}^{N_p} \left( \sum_{x=1}^{N} \hat{c}^{\dagger}_{x}P_{xn}^{\phantom{\dagger}}\right) | 0 \rangle\;.
\end{equation}
The trial wave function is thus defined by the rectangular ${N\times N_p}$ matrix $\mathbf{P}$ with $N_p$ the number of particles. Given this trial wave function, and assuming that it is non-orthogonal to the ground state, observables can be obtained from
\begin{equation}
	\langle \hat{O} \rangle_0 = \lim_{\Theta \rightarrow \infty} \frac{ \langle \Psi_\text{T} | {\rm e}^{-\Theta \hat{H}} \hat{O}\, {\rm e}^{-\Theta \hat{H}} | \Psi_\text{T}\rangle } { \langle \Psi_\text{T} | {\rm e}^{-2 \Theta \hat{H}} | \Psi_\text{T}\rangle} \;,
\end{equation}
for large but finite values of the projection parameter $\Theta$. To compute the imaginary time propagation one first discretizes the imaginary time, ${L_{\Theta} \Delta \tau = 2 \Theta}$, and then carries out a Trotter decomposition to isolate the Hubbard interaction term,
\begin{equation}
	  {\rm e}^{-2 \Theta \hat{H}} = \prod_{\tau=1}^{L_{\Theta}} {\rm e}^{- \Delta \tau \hat{H}_T/2}  {\rm e}^{- \Delta \tau \hat{H}_U} {\rm e}^{- \Delta \tau \hat{H}_T/2} + \mathcal{O}(\Delta\tau^2)\;.
\end{equation} 
Hereafter we neglect the systematic and controllable Trotter error \cite{Rost12}. The key point of the algorithm is to use a Hubbard-Stratonovitch (HS) transformation to reformulate the many-body imaginary time propagator as a sum of one-body problems by introducing an auxiliary field. We have adopted the discrete decomposition \cite{Hirsch83}, 
\begin{equation}
	{\rm e}^{-\Delta\tau \frac{U}{2} \left(\hat{n}_{\pmb i} -1 \right)^2 } = \gamma \sum_{s=\pm 1} {\rm e}^{s \alpha \left(n_{{\pmb i}\uparrow} - n_{{\pmb i}\downarrow} \right)}\;,
\end{equation}
with $ \cosh(\alpha) = {\rm e}^{\Delta \tau U/2}$ and $\gamma = \frac{1}{2}{\rm e}^{-\Delta \tau U /2}$. With this transformation, the imaginary time propagation reads
\begin{equation}
	 {\rm e}^{-2 \Theta \hat{H}} = \gamma^{N L_{\Theta}} \sum_{ {\pmb s}_{1} \cdots{\pmb s}_{L_{\Theta}} } \prod_{\tau=1}^{L_{\Theta}} {\rm e}^{\hat{\mathbf{c}}^{\dagger} \mathbf{A}({\pmb s}_{\tau})\hat{\mathbf{c}} } {\rm e}^{ - \Delta \tau \hat{\mathbf{c}}^{\dagger} \mathbf{T} \hat{\mathbf{c}} }\;,
\end{equation}
with $ A({\pmb s}_{\tau})_{xy} = \delta_{xy}\,s_{{\pmb i}\tau}\,\sigma\,\alpha$. Recall that $x=({\pmb i}\sigma)$ and that $\sigma$ takes the value $1$ ($-1$) for the up (down) $z$-component of the spin. For a given configuration of Ising variables, ${\pmb s} \equiv \left\{ {\pmb s}_{1} \cdots {\pmb s}_{L_{\Theta}} \right\}$, we now have to solve a free fermion problem in an external space and time dependent field. Since under a single body propagator a Slater determinant remains a Slater determinant, 
\begin{equation}
	{\rm e}^{\hat{\mathbf{c}}^{\dagger} \mathbf{h} \hat{\mathbf{c}}}\prod_{n=1}^{N_p} \left( \hat{\mathbf{c}}^{\dagger} \mathbf{P}\right)_n | 0 \rangle = \prod_{n=1}^{N_p} \left( \hat{\mathbf{c}}^{\dagger} {\rm e}^\mathbf{h} \mathbf{P}\right)_n | 0 \rangle\;, 
\end{equation}
and the overlap of two Slater determinants defined by $\mathbf{P}$ and $\mathbf{P}'$ is a determinant,
\begin{equation}
   \langle \Psi'_\text{T} | \Psi_\text{T}\rangle = \det\left( \mathbf{P}'^{\dagger} \mathbf{P} \right)\,,
\end{equation}
we can integrate out the fermions to obtain: 
\begin{equation}
   \langle \hat{O} \rangle_0 = \sum_{\pmb{s}} P_{\pmb s} \langle \hat{O} \rangle_{\pmb s}\;.
\end{equation}
Here, 
\begin{equation}
	P_{\pmb s} = \frac{ \det \left( \mathbf{U}^{<}_{\pmb s} \mathbf{U}^{>}_{\pmb s} \right) } 
      { \sum_{\pmb{s}} \det \left( \mathbf{U}^{<}_{\pmb s} \mathbf{U}^{>}_{\pmb s} \right) }\;, 
\end{equation}
with
\begin{eqnarray*}
   \mathbf{U}^{>}_{\pmb s} & = & \left( \prod_{\tau=1}^{L_{\Theta}/2} {\rm e}^{ \mathbf{A}( {\pmb s}_{\tau})} {\rm e}^{ - \Delta \tau \mathbf{T} } \right) \mathbf{P}\;, \\
   \mathbf{U}^{<}_{\pmb s} & = & \mathbf{P}^{\dagger }\left( \prod_{\tau=L_{\Theta}/2}^{L_{\Theta}} {\rm e}^{ \mathbf{A}( {\pmb s}_{\tau})} {\rm e}^{ - \Delta \tau \mathbf{T} } \right) \;.
\end{eqnarray*}
For a single Ising field configuration ${\pmb s}$, Wick's theorem holds, such that it suffices to compute the single-particle Green function, 
\begin{equation}
   \label{Green.eq}
   G_{xy}(\pmb s) \equiv \langle \hat{c}^{\dagger}_x \hat{c}^{\phantom{\dagger}}_y \rangle_{\pmb s} = \left[\mathbf{U}^{>}_{\pmb s} \left(\mathbf{U}^{<}_{\pmb s}\mathbf{U}^{>}_{\pmb s} \right)^{-1}\mathbf{U}^{<}_{\pmb s} \right]_{yx}\;,
\end{equation}
from which arbitrary observables can be computed. Finally the sum over the fields will be carried out with Monte Carlo importance sampling. This completes our brief review of the PQMC auxiliary field algorithm. For further details, we refer the reader to Ref.~\cite{Assaad08_rev}.

Alternatively, we can recast the above in terms of a density matrix,
\begin{equation}
\label{Tarun.eq}
	\hat{\rho} \equiv | \Psi_0 \rangle \langle \Psi_0 | = \sum_{\pmb s } P_{\pmb s} \; \hat{\rho} (\pmb s)\;,
\end{equation}
with 
\begin{equation*}
	\hat{\rho}(\pmb s) = \det\left[ \mathbf{1} - \mathbf{G}(\pmb s) \right]\, {\rm e}^{ -\hat{\mathbf{c}}^{\dagger} \ln\left[ \mathbf{G}^{-1}(\pmb s) - \mathbf{1}\right]\,\hat{\mathbf{c}} } \;,
\end{equation*}
such that 
\begin{equation}
   \langle \hat{O} \rangle_0 = {\rm Tr} \left[ \hat{\rho} \; \hat{O} \right].
\end{equation}
The above is essentially the central result of Ref.~\onlinecite{Grover13} and is a direct consequence of the validity of Wick's theorem for a given configuration of HS fields, as well as the formulation of the density matrix for a Gaussian problem \cite{Peschel03}
\footnote{Note that $\hat{\rho} (\pmb s) $ is not a density matrix but a general Gaussian  operator.  In particular,  
${\rm Tr} \left[ \hat{\rho} (\pmb s)  \hat{c}^{\dagger}_{x} \hat{c}_{x} \right]   \equiv G_{x,x}(\pmb s) $ can become negative or even complex depending upon the choice of the HS transformation.}.
Starting from Eq.~(\ref{Tarun.eq}), the reduced density matrix obtained by splitting the Hilbert space as ${\cal H} = {\cal H}_\text{A} \otimes{\cal H}_\text{B} $ and tracing over ${\cal H}_\text{B}$, reads:
\begin{equation}
   \label{Tarun1.eq}
	{\hat \rho}_\text{A} \equiv {\rm Tr}_{ {\cal H}_\text{B}} | \Psi_0 \rangle \langle \Psi_0 | = \sum_{\pmb s } P_{\pmb s} \; \hat{\rho}_\text{A}(\pmb s)\;,
\end{equation}
with 
\begin{equation*}
	\hat{\rho}_\text{A}({\pmb s}) = \det\left[ \mathbf{1} - \mathbf{G}_\text{A}(\pmb s) \right] {\rm e}^{ -\hat{\pmb a}^{\dagger} \ln\left[ \mathbf{G}_\text{A}^{-1}(\pmb s) - \mathbf{1}\right]\hat{\pmb a} }\;.
\end{equation*}
Here $\hat{\mathbf{c}}^{\dagger} = \left(\hat{\pmb a}^{\dagger}, \hat{\pmb b}^{\dagger} \right) $ and ${\left[\mathbf{G}_\text{A} (\pmb s)\right]_{zz'} = \langle \hat{a}^{\dagger}_{z} \hat{a}_{z'} \rangle_{\pmb s}}$ is the Green function restricted to subsystem ${\cal H}_\text{A}$ \footnote{With this definition of ${\hat \rho_\text{A}}$ one will readily verify that for any observable $\hat{A} \in {\cal H}_\text{A}$, ${\rm Tr }_\text{A} \left[\hat{\rho}_{{\cal H}_\text{A}} \hat{A} \right] = {\rm Tr } \left[\hat{\rho} \hat{A} \right]$}.

%------------------------------------------------------------------------------------------------------------
\subsection{Numerical stabilization for Renyi entropies}
%------------------------------------------------------------------------------------------------------------

\begin{figure}
   \includegraphics[width=\linewidth]{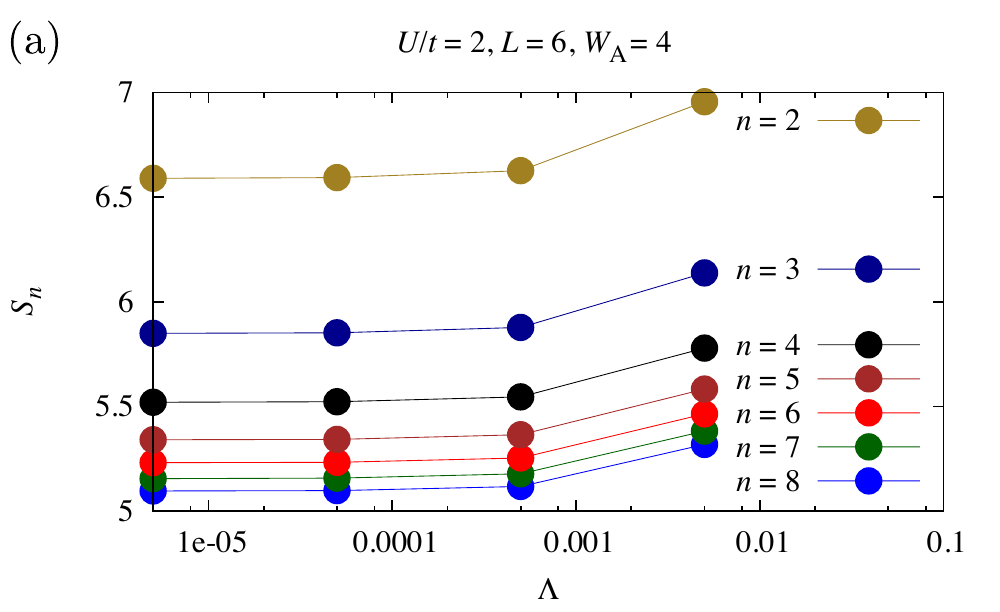}\\
   \includegraphics[width=\linewidth]{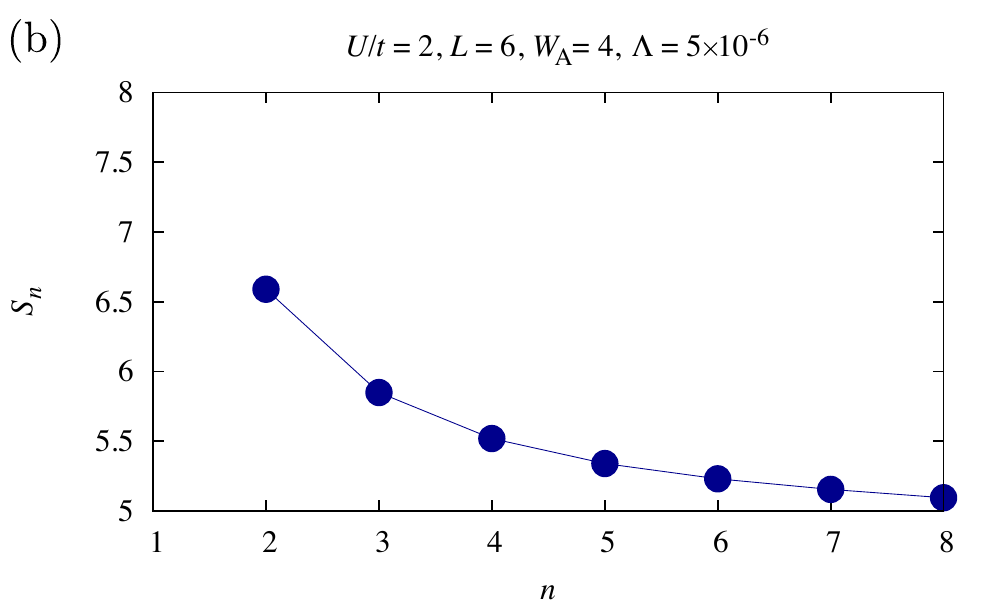}
   \caption{Renyi entanglement entropies as a function of the regularization parameter $\Lambda$ (a) and order $n$ (b) for a $6 \times 6$ Kane-Mele Hubbard model at ${U/t = 2}$, ${\lambda/t = 0.2}$ and ${t'/t = 1}$ and the region A defined in Fig.~\ref{Latt.fig}. Errorbars are smaller than the symbol size. For these simulations, we have used a projection parameter  $\Theta t =40$. This guarantees  convergence within our error-bars. For each replica, we have carried out $500 \times 10^3$ sweeps. Each sweep consists of sequentially updating each Ising spin in the space-imaginary time lattice.}
   \label{Renyi_Test.fig}
\end{figure}

The $n^{th}$ Renyi entropy is defined as:
\begin{equation}
   S_n = -\frac{1}{n-1} \ln {\rm Tr}_{{\cal H}_\text{A}} \left[ {\hat \rho}_\text{A}^n \right]\;.
\end{equation}
To evaluate the above quantity, one introduces $n$ replicas, (or $n$ independent QMC simulations) such that 
\begin{eqnarray}
\label{Renyi.Eq}
	{\rm e}^{-(n-1)S_n} 
	   & = & \sum_{{\pmb s}^1, \cdots, {\pmb s}^n } P_{{\pmb s}^n} \cdots P_{{\pmb s}^1} {\rm Tr }_{{\cal H}_\text{A}} 
            \left[ \hat{\rho}_\text{A}({\pmb s}^n) \cdots \hat{\rho}_\text{A}({\pmb s}^1) \right] \nonumber\\
      & = & \sum_{{\pmb s}^1, \cdots, {\pmb s}^n } P_{{\pmb s}^n} \cdots P_{{\pmb s}^1} \prod_{m=1}^{n} \det\left[ \mathbf{1} - \mathbf{G}_\text{A}({\pmb s}^m) \right] \nonumber\\
      &   & \quad\quad\quad\times \det \left[ \mathbf{1} + \prod_{m=1}^{n} \frac{\mathbf{G}_\text{A}({\pmb s}^m)}{\mathbf{1} - \mathbf{G}_\text{A}({\pmb s}^m) } \right] \;.
\end{eqnarray}
Here we have made use of the identity: ${{\rm Tr} \left[{\rm e}^{ \hat{\pmb c }^\dagger \mathbf{T}_1 \hat{\pmb c}} \cdots {\rm e}^{ \hat{\pmb c }^\dagger \mathbf{T}_n \hat{\pmb c}}\right] = \det \left[ \mathbf{1} + {\rm e}^{\mathbf{T}_1} \cdots {\rm e}^{\mathbf{T}_n} \right]}$ and the superscripts denote the replicas. At $n=2$ one only needs the knowledge of $\mathbf{G}_\text{A}({\pmb s})$ to compute the Renyi entropy and one recovers Eq.~(6) of Ref.~\cite{Grover13}.   For $n>2$ we were not able to avoid an explicit calculation of the inverse of $\mathbf{G}_\text{A}({\pmb s})$ which will also be required  to compute the entanglement spectrum. However, $\mathbf{G}_\text{A}({\pmb s})$  is in general a singular matrix. To show this, let us first consider the Green function in Eq.~(\ref{Green.eq}). As pointed out in \cite{Feldbach00} the Green function in the PQMC algorithm is a projector operator,
\begin{equation}
   \mathbf{G}^2({\pmb s}) = \mathbf{G}({\pmb s}) \;,
\end{equation}
such that the eigenvalues of $\mathbf{G}({\pmb s})$ are given by $0$ or $1$. Since the $ {\rm Tr} \left[\mathbf{G}({\pmb s}) \right] = N_p$, $\mathbf{G}({\pmb s})$ contains $N_p$ non-vanishing and $N-N_p$ vanishing eigenvalues. Thereby ${\mathbf{G}^{-1}({\pmb s})}$, and by the same token ${\left[\mathbf{1} - \mathbf{G}({\pmb s}) \right]^{-1}}$, does not exist. In the absence of interactions, the above is merely stating that in the zero temperature limit the occupation number of  single-particle eigenstates is given by $1$ or $0$. That ${\mathbf{G}^{-1}({\pmb s})}$ does not exist does not necessarily mean that ${\mathbf{G}_\text{A}({\pmb s})}$ is singular. In fact for the half-filled case, if the domain A (cf. Fig.~\ref{Latt.fig}) contains $\sqrt{N_s}$ out of $N_s$ sites, ${\mathbf{G}_\text{A}({\pmb s})}$ turns out to be generically regular. On the other hand, if the domain A contains more sites than the number of particles $N_p$, $\mathbf{G}_\text{A}({\pmb s})$ is indeed singular.

At finite temperatures the Green function matrix is never singular due to thermal broadening. This suggests the following regularization scheme. Let $\mathbf{G}_{\perp}({\pmb s})$ be a projector on the vector space spanned by the elements of the kernel of $\mathbf{G}({\pmb s})$. One can then consider the quantity
\begin{equation}
	\tilde{\mathbf{G}}({\pmb s}) = ( \mathbf{1} - \Lambda) \mathbf{G}({\pmb s}) + \Lambda \mathbf{G}_{\perp}({\pmb s})\;. 
\end{equation}
Essentially, $\Lambda$ introduces a finite {\it temperature} since the occupation number of occupied (unoccupied) single-particle states is reduced (enhanced) from 1 (0) to $1 - \Lambda$ ($\Lambda$) \footnote{Note that for the here considered half-filled case where $N_p = N/2$, this regularization scheme does not alter the particle number.}. By definition $\tilde{\mathbf{G}}({\pmb s})$ is regular. From $\tilde{\mathbf{G}}({\pmb s})$ we can construct $\tilde{\mathbf{G}}_\text{A}({\pmb s})$ which equally proves to be regular. If $\Lambda$ is small, we do not expect this regularization scheme to alter the nature of the entanglement spectrum, nor the entanglement entropies. Furthermore, within the same QMC run, it is possible to extrapolate to $\Lambda \rightarrow 0$.   

We have tested the above for the Kane-Mele Hubbard model. Figure~\ref{Renyi_Test.fig}(a) shows the Renyi entanglement entropies up to ${n = 8}$ on a $6\times 6 $ lattice at ${U/t =2}$, ${\lambda/t = 0.2}$ and ${t=t'}$ with region A being a ribbon of width $W_\text{A}=4$ (see Fig.~\ref{Latt.fig}). For this specific problem, values of $\Lambda = 5\times 10^{-6} $ suffice to guarantee stability and to render systematic errors negligible within error bars. From now on we will always use the presented regularization scheme and generically use the notation $\mathbf{G}_\text{A}({\pmb s})$ instead of $\tilde{\mathbf{G}}_\text{A}({\pmb s})$.   

Instead of implementing the above regularization scheme, one could use a finite temperature algorithm \cite{White89,Assaad08_rev}. There are  however many cases in which an adequate choice of the trial wave function leads to considerable computational gain insofar as ground state properties are concerned \cite{Capponi00}. In this case, the above regularization scheme  will certainly be more useful than a finite-temperature simulation.

Figure~\ref{Renyi_Test.fig}(b) plots the Renyi entropies as a function of $n$. It is interesting to note that extrapolation to ${n=1}$, corresponding to the von Neumann entropy, is difficult. Detailed knowledge of the higher order Renyi entropies has been used in Ref.~\cite{Chung13} to access the entanglement spectra. In the next section we will present an alternative approach.

%============================================================================================================
\section{Accessing the entanglement spectrum}\label{ES.sec}
%============================================================================================================
%
\begin{figure}
   \includegraphics[width=\linewidth]{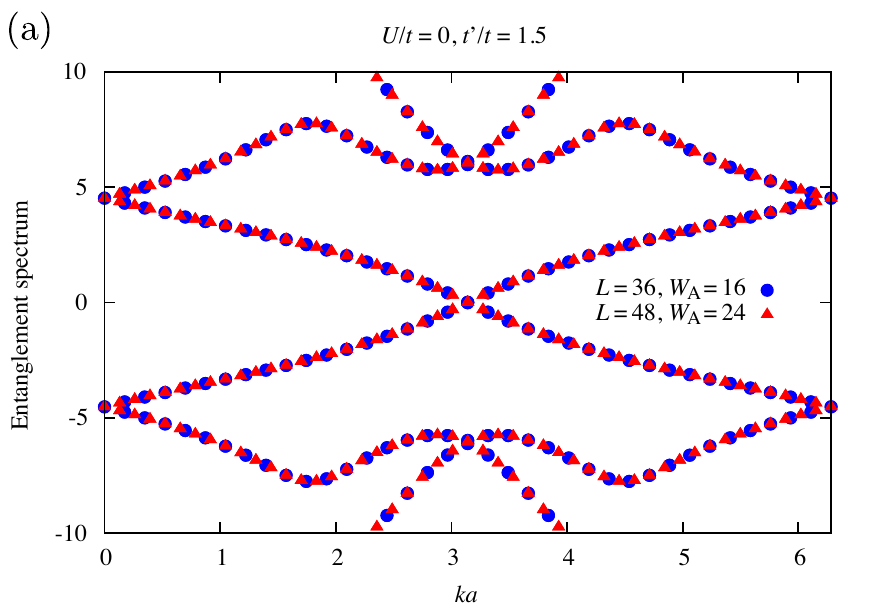} \\
   \includegraphics[width=\linewidth]{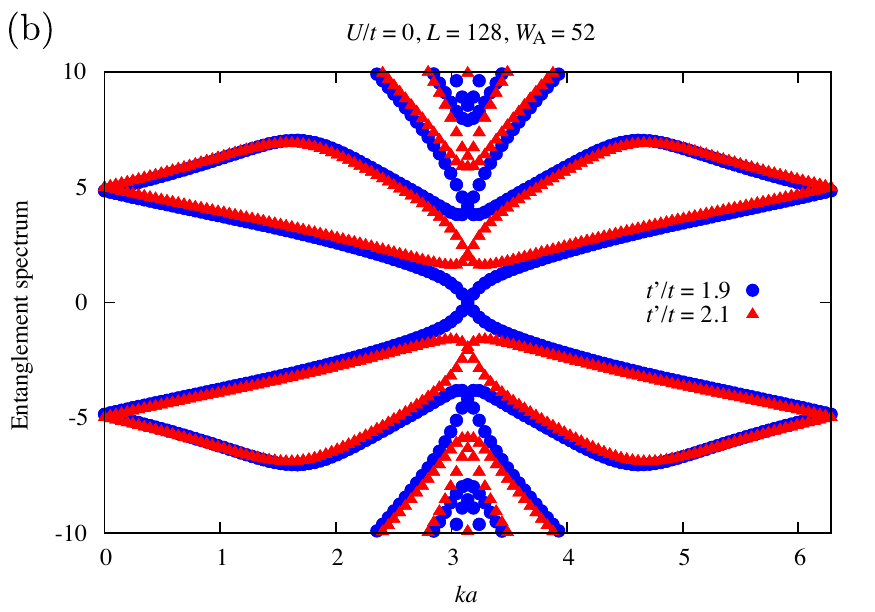} \\
   \includegraphics[width=\linewidth]{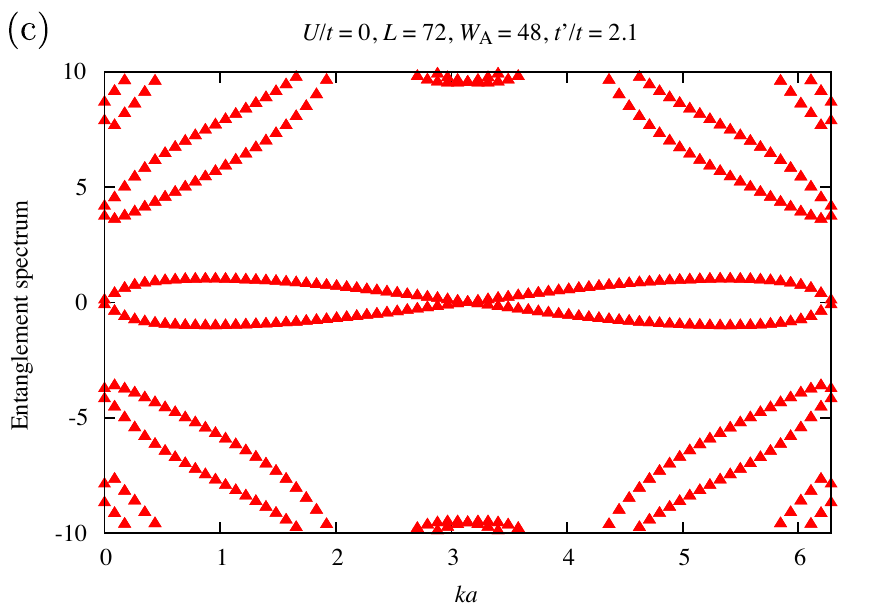}
   \caption{Entanglement spectrum of the dimerized Kane-Mele model at ${\lambda/t=0.2}$ and ${U/t=0}$. In (a) and (b) the dimerization runs along the ${\pmb \delta}_1$ direction (See Fig.~\ref{Latt.fig}). In (c) the dimerization runs along the ${\pmb \delta}_2$ direction. The spectrum has been obtained by inverting eq.~(\ref{nonint}).}
\label{EntanglementU0.fig}
\end{figure}
Generically, in quantum Monte Carlo simulations, we can access the spectrum of the Hamiltonian by computing imaginary time displaced correlation functions in various channels. Similarly, for the entanglement Hamiltonian, given by 
\begin{equation}
	\hat{\rho}_\text{A} \equiv \frac{{\rm e}^ {-\hat{H}_\text{E}}} {{\rm Tr }_{{\cal H}_\text{A}} \left[ {\rm e}^ {-\hat{H}_\text{E} } \right]}\;,
\end{equation}
we can define the following {\it replica time} displaced correlation functions
\begin{eqnarray}
\label{replica.eq}
   S^{E}_{O}(\tau_\text{E})
      & \equiv & \langle \hat{O}^{\dagger} (\tau_\text{E}) \hat{O} \rangle_\text{A} \nonumber\\
      & \equiv & \frac{{\rm Tr }_{{\cal H}_\text{A}} \left[ {\rm e}^ {-(n-\tau_\text{E}) \hat{H}_\text{E} } \hat{O}^{\dagger} {\rm e}^ {-\tau_\text{E} \hat{H}_\text{E} } \hat{O} \right]} { {\rm Tr }_{{\cal H}_\text{A}} \left[ {\rm e}^ {- n \hat{H}_\text{E} } \right] }\;,
\end{eqnarray}
for an observable ${O \in {\cal H}_\text{A}}$. Here $\tau_\text{E}$ and $n$ are integers with ${\tau_\text{E} < n}$. In analogy to imaginary time, one can insert a complete set of eigenstates of $\hat{H}_\text{E}$, ${\hat{H}_E |\Psi_i\rangle =  E_i |\Psi_i \rangle}$ to obtain
\begin{subequations}
\begin{equation}
\label{Analytical_continuation.eq}
   S^{E}_{O}(\tau_\text{E}) = \frac{1}{\pi} \int {\rm d} \omega  \frac{{\rm e}^{- \tau_\text{E} \omega} }{ 1 \pm  {\rm e}^{-\tau_\text{E}  \omega} } A^{E}_{O}(\omega) \;,
\end{equation}
with
\begin{eqnarray}
	A^{E}_{O}(\omega) & = &\frac{\pi}{\sum_{i} {\rm e}^{- n E_i} } \sum_{i,j} {\rm e}^{- n E_i} |\langle \Psi_i | \hat{O} | \Psi_j \rangle |^2 \nonumber\\
   && \times \;\delta ( \omega + E_i - E_j) \left( 1 \pm {\rm e}^{-\beta \omega} \right) \;.
\end{eqnarray}
\end{subequations}
Here, the plus (minus) sign refers to single-particle (particle-hole) excitations. Hence, in principle entanglement spectral functions can be computed by using standard maximum entropy methods \cite{Jarrell96,Sandvik98,Beach04a}. Note however, that in contrast to imaginary time, replica time is quantized to integer values in our approach. Hence, with the use of the maximum entropy method we will only be able to resolve the low lying spectral features. 

The Monte Carlo evaluation of Eq.~(\ref{replica.eq}) is very similar to computing imaginary time displaced correlation functions in the realm of the finite temperature formulation of the auxiliary field determinant QMC algorithm \cite{Assaad08_rev}. To highlight the similarity we will introduce the notation: ${\hat{B}_\text{A}({\pmb s}) = {\rm e}^{ -\hat{\pmb a}^{\dagger} \ln\left[ \mathbf{G}_\text{A}^{-1}(\pmb s) - \mathbf{1}\right]\hat{\pmb a} }}$, and ${\mathbf{B}_\text{A}({\pmb s}) = {\rm e}^{ - \ln\left[ \mathbf{G}_\text{A}^{-1}(\pmb s) - \mathbf{1}\right] }}$, such that
\begin{widetext}
\begin{subequations}
\begin{equation}
\label{Time_displaced.eq}
   \langle \hat{O}^{\dagger} (\tau_\text{E}) \hat{O} \rangle_\text{A} = \frac{ \sum_{{\pmb s}^1, \cdots, {\pmb s}^n } P_{{\pmb s}^n} \cdots P_{{\pmb s}^1} Z_n\left( {\pmb s^n} \cdots {\pmb s^1} \right) \langle \hat{O}^{\dagger} (\tau_\text{E}) \hat{O} \rangle_\text{A}\left( {\pmb s}^n \cdots {\pmb s}^1 \right) } 
   { \sum_{{\pmb s}^1, \cdots, {\pmb s}^n } P_{{\pmb s}^n} \cdots P_{{\pmb s}^1} Z_n\left( {\pmb s^n} \cdots {\pmb s^1} \right) } \;,
\end{equation}
with
\begin{equation}
%\label{Time_displaced2.eq}
	Z_n\left( {\pmb s^n} \cdots {\pmb s^1} \right) = \left[ \prod_{m=1}^{n} \det\left[ \mathbf{1} - \mathbf{G}_\text{A}({\pmb s}^m) \right] \right]
         {\rm Tr}_{ {\cal H}_\text{A}}   \left[ \hat{B}_\text{A}({\pmb s}^{n}) \cdots   \hat{B}_\text{A}({\pmb s}^{1}) \right] \;,
\end{equation}
%\det \left[ \mathbf{1} + \prod_{m=1}^{n} \mathbf{B}_\text{A}({\pmb s}^{m}) \right]
and
\begin{equation}
   \label{Time_displaced2.eq}
	\langle \hat{O}^{\dagger} (\tau_\text{E}) \hat{O} \rangle_\text{A}\left( {\pmb s}^n \cdots {\pmb s}^1 \right) = 
        \frac{ {\rm Tr}_{ {\cal H}_\text{A}} \left[ \hat{B}_\text{A}({\pmb s}^{n}) \cdots \hat{B}_\text{A}({\pmb s}^{\tau_\text{E}+1} ) \hat{O}^{\dagger}\hat{B}_\text{A}({\pmb s}^{\tau_\text{E}}) \cdots \hat{B}_\text{A}({\pmb s}^{1}) \hat{O}\right] }  
    { {\rm Tr}_{ {\cal H}_\text{A}}   \left[ \hat{B}_\text{A}({\pmb s}^{n}) \cdots   \hat{B}_\text{A}({\pmb s}^{1}) \right] }\;.
\end{equation}
\end{subequations}
\end{widetext}
The denominator in Eq.(\ref{Time_displaced.eq}) is merely ${\rm e}^{-(n-1)S_n}$. To compute the numerator one has to evaluate the replica time displaced correlation function of Eq.~(\ref{Time_displaced2.eq}). Technically speaking this is identical to computing imaginary time displaced correlation functions in the realm of the finite temperature auxiliary field QMC algorithm. For a given value of the HS fields Wick's theorem holds such that it suffices to compute the single-particle replica time displaced Green function. We refer the reader to Ref.~\cite{Assaad08_rev} for a detailed description. To implement the above, we need to run $n$ independent PQMC simulations (or replicas) \footnote{Alternatively, replica time displaced Green functions can be obtained in a single run from measurements which are sufficiently separated in imaginary time, such as to guarantee statistical independence.}. At any one time, the configurations will be distributed according to the probability distribution $P_{{\pmb s}^n} \cdots P_{{\pmb s}^1} $. With this configuration of HS fields, we compute the $\mathbf{B}_\text{A}({\pmb s}^m)$ and $\mathbf{G}_\text{A}({\pmb s}^m)$ matrices from which can evaluate the n-th Renyi entropy as well as the replica time displaced correlation functions \footnote{For the Kane-Mele Hubbard model at half-band filling, one can show that ${Z_n\left( {\pmb s^n} \cdots {\pmb s^1} \right)}$ is positive. Hence instead of using a re-weighting scheme, one can in principle devise an algorithm which directly samples $P_{{\pmb s}^n} \cdots P_{{\pmb s}^1} Z_n\left( {\pmb s^n} \cdots {\pmb s^1} \right) $ }.

We have tested the approach for the Kane-Mele Hubbard model defined in Eq.~(\ref{Hamiltonian.eq}). In the non-interacting case, the entanglement Hamiltonian satisfies the relation
\begin{equation}
	\frac{\mathbf{1}}{2} - \mathbf{G}_\text{A} = \frac{1}{2} \tanh \left( \mathbf{H}_\text{E}/2 \right)\;.
\label{nonint}
\end{equation} 
Using the {\it spectral flattening trick} \cite{Kitaev06,Fidkowski10,Turner10} we can adiabatically connect $(\frac{\mathbf{1}}{2} - \mathbf{G})$ to the original Hamiltonian. Thereby both Hamiltonians share the same topological character and restricting $(\frac{\mathbf{1}}{2} - \mathbf{G})$ to a region A with edges will reveal the edge physics of the original Hamiltonian. The eigenvalues of the matrix $(\frac{\mathbf{1}}{2} - \mathbf{G})$ take values $\pm 1/2$. Thereby the eigenvalues of $(\frac{\mathbf{1}}{2} - \mathbf{G}_\text{A})$ are restricted to the range $(-1/2; 1/2)$ and the spectrum of $H_\text{E} $ is not bounded. In the plots of Fig.~\ref{EntanglementU0.fig} we consider an energy window $[-10t,10t]$. Since the considered cut A preserves translation symmetry along the $\mathbf{a}_1$ direction, ${ka = \mathbf{k}\cdot\mathbf{a}_1}$ remains a good quantum number. Alternative choices of A which do not break any lattice symmetries have been proposed \cite{Legner13}.

In the absence of interactions, the critical dimerization where the topological phase transition between topological and trivial band insulators takes place reads $t'_c/t = 2$. As shown in Fig.~\ref{EntanglementU0.fig}(a), at ${t'/t = 1.5 < t'_c/t}$ one observes the typical edge signature of the quantum spin Hall (QSH) state: a single crossing at the time-reversal symmetric point ${ka = \pi}$, which due to particle-hole symmetry, is pinned at zero energy. In the vicinity of the critical point, the velocity of the edge state grows, and beyond $t'_c/t$ a gap opens [cf. see Fig.~\ref{EntanglementU0.fig}(b)]. 

It is interesting to note that the entanglement spectrum depends on the direction in which the dimerization is imposed. In particular, for a dimerization along the ${\pmb \delta }_2 $ bond, the entanglement spectrum is given in Fig.~\ref{EntanglementU0.fig}(c). For ${t'> t'_c}$ this is still the spectrum of a trivial insulator: the surface state is disconnected from the bulk and an even number of crossings at $ka = \pi$ and 0 are visible \cite{Hasan10_rev}. Hence, care has to be taken when analyzing the entanglement spectrum before drawing conclusions on the nature of the state.

\begin{figure*}
\includegraphics[width=0.48\linewidth]{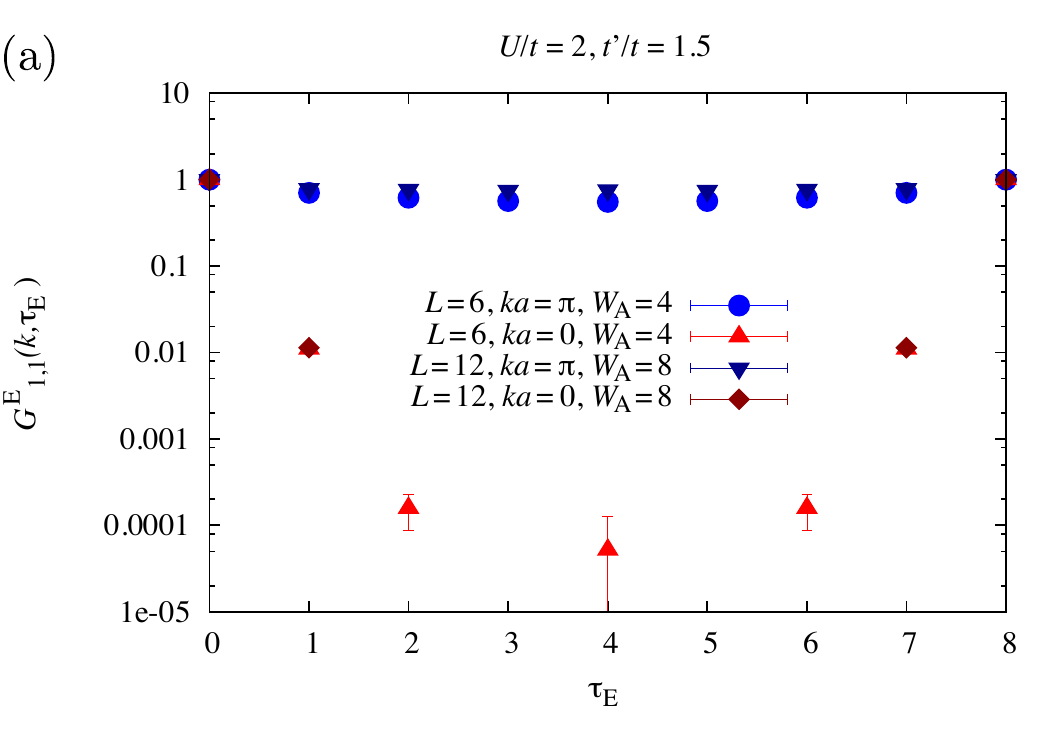}\includegraphics[width=0.25\linewidth]{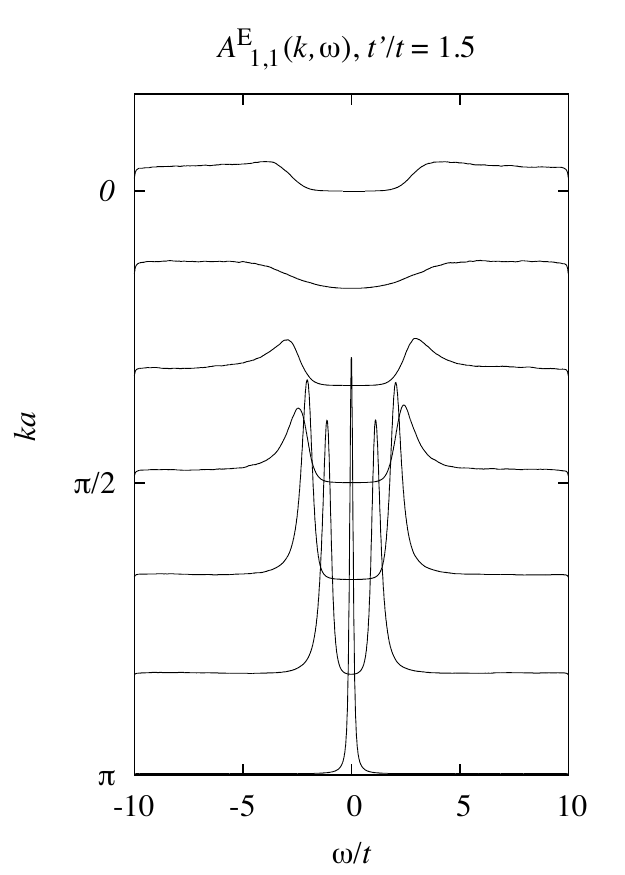}\\
\includegraphics[width=0.48\linewidth]{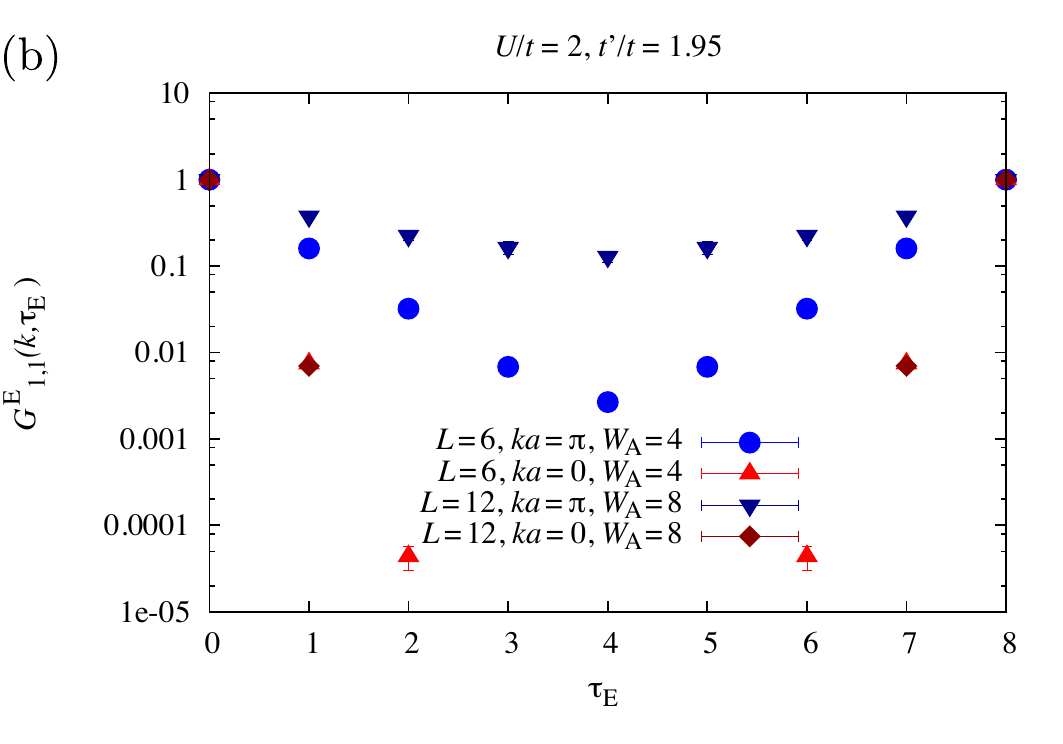}\includegraphics[width=0.25\linewidth]{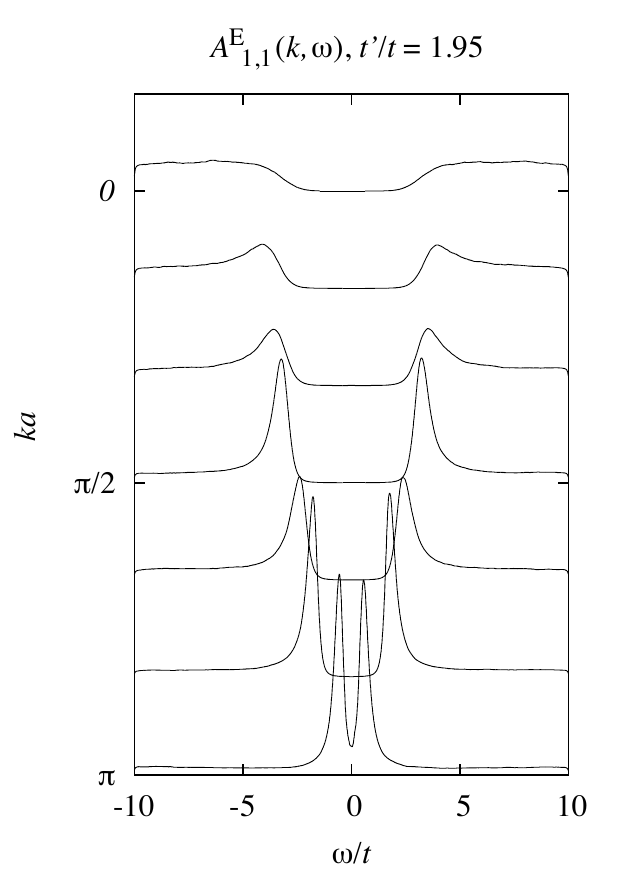}\\
\includegraphics[width=0.48\linewidth]{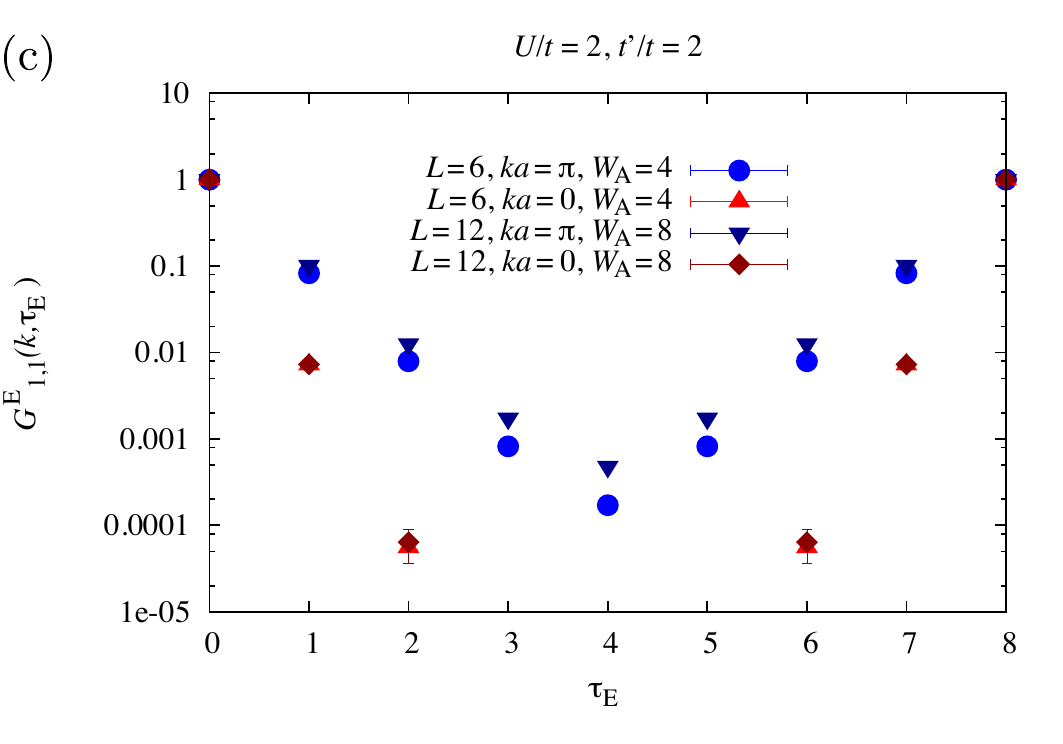}\includegraphics[width=0.25\linewidth]{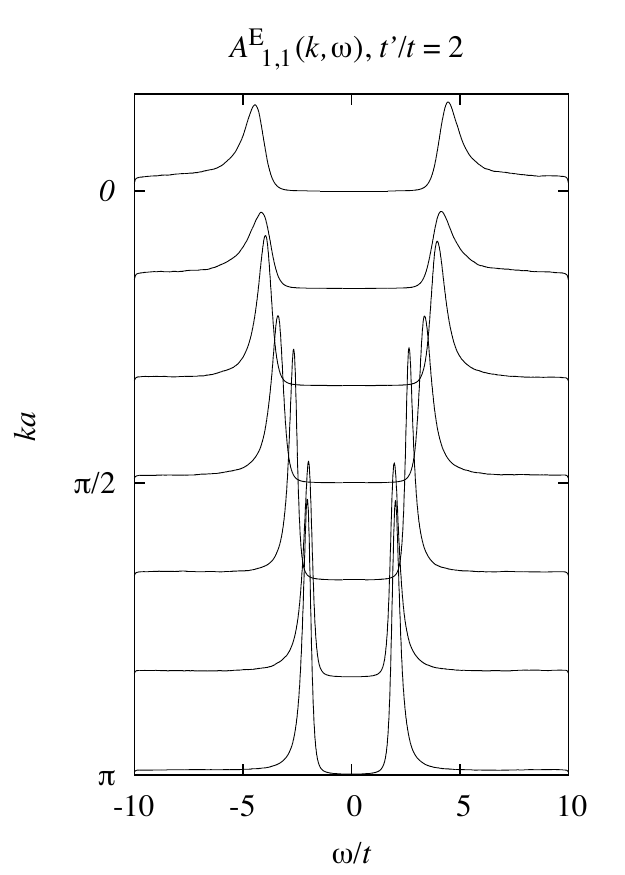}
\caption{Replica time displaced single-particle Green function (left) and the corresponding single-particle spectral function (right) for the Kane-Mele Hubbard model at ${U/t=2}$, ${\lambda/t = 0.2}$ and different values of the dimerization ${t'/t}$ across the transition from the QSH insulator to the topologically trivial band insulator. In computing the single-particle spectral function we have considered the ${12\times 12}$ lattice with ${W_\text{A} = 8}$. For the largest system sizes, ${L=12}$, we have used  ${40 \times 10^3}$ sweeps per replica and set the imaginary time propagation  parameter to ${\Theta t = 40}$. The spectral functions are obtained with the stochastic maximum entropy method \cite{Sandvik98,Beach04a}.}
\label{EntanglementU2_Spectral.fig}
\end{figure*}

In the presence of interactions, we can compute the replica time displaced single-particle Green function
\begin{equation}
	G^\text{E}_{\alpha\beta}(k,\tau_\text{E}) = \sum_{\sigma}\langle \hat{a}_{k\alpha\sigma}^{\dagger}(\tau_\text{E})\, \hat{a}_{k\beta\sigma} \rangle_\text{A}\;,
\end{equation}
where the indices $\alpha,\beta$ run over the sites across the width of the ribbon defined by region A, and $k$ is the conserved crystal moment associated with translation invariance along the $\mathbf{a}_1$ direction. As mentioned above, $G^\text{E}_{\alpha\beta}(k,\tau_\text{E})$, together with the replica time displaced correlation functions of other observables, provides the input for maximum entropy methods which in principle, allow us to determine the spectrum of $\hat{H}_\text{E}$. On the left side of Fig.~\ref{EntanglementU2_Spectral.fig} we show $G^\text{E}_{\alpha\beta}(k,\tau_\text{E})$ at $U/t = 2$ and for various values of the dimerization. We consider the orbital $\alpha=1$ situated on the edge of the ribbon as well as two momenta, ${ka = \pi}$ and 0. At ${t'/t = 1.5}$, $G^{E}_{1,1}(\pi,\tau_\text{E})$ does not decay as a function of replica time $\tau_\text{E}$ thus signaling a gapless single-particle excitation at $ka = \pi$. In contrast, at ${k=0}$ a clear gap is visible. We hence conclude that we are in a topological phase. At ${t'/t = 1.95}$ analysis of the same quantity shows large finite size effects, which one can interpret in terms of the vanishing of the single-particle gap at ${ka = \pi}$, and a robust gap at ${k=0}$. Hence the data at ${t'/t = 1.95}$ allows interpretation in terms of a topological phase. On the other hand, at ${t'/t = 2}$ analysis of the data as a function of system size shows that the gap remains robust at both considered $k$-points. Based on the entanglement spectra of the single-particle Green function one can conclude that the topological phase transition occurs in the interval ${1.95 < t'/t < 2}$. This is in accordance with independent results of Ref.~\cite{Lang13_1}.

Using the maximum entropy method and Eq.~(\ref{Analytical_continuation.eq}) we can analytically continue the replica time single-particle spectral function to obtain the {\it entanglement} single-particle spectral function, $A^\text{E}_{1,1}(\pi,\omega)$. This quantity is plotted on the right side of Fig.~\ref{EntanglementU2_Spectral.fig}. The analytical continuation confirms the conclusions drawn from the replica time data. Deep in the topological phase at $t'/t = 1.5$ [Fig.~\ref{EntanglementU2_Spectral.fig}(a)] a gapless edge mode is visible. In the trivial band insulating phase at $t'/t = 2$ [Fig.~\ref{EntanglementU2_Spectral.fig}(c)] the edge mode acquires a gap. Close to the transition at $t'/t = 1.95$ [Fig.~\ref{EntanglementU2_Spectral.fig}(b)] we interpret the gap at $k=\pi$ in terms of a finite size effect. 

%============================================================================================================
\section{Conclusions}\label{Conclusions.sec}
%============================================================================================================

Based on the work of T. Grover \cite{Grover13} we have demonstrated how to compute in a numerically stable manner Renyi entanglement entropies in the realm of the projective auxiliary field quantum Monte Carlo algorithm. Furthermore we have introduced replica time displaced correlation functions from which the low-lying properties of the entanglement spectrum can be reconstructed. The strong point of such an approach is two-fold: One can study the entanglement spectrum in different channels, single-particle, particle-hole, or particle-particle, and one can classify the spectrum in terms of symmetries allowed by the choice of the subsystem A. We have successfully tested this approach for the topological quantum phase transition from the QSH insulator to a trivial band insulator in the weak coupling regime around ${U/t = 2}$. The weak point of our present implementation is that the replica time displaced correlation functions become statistically noisy at stronger couplings. This originates from the fact that the stochastic evaluation of the Renyi entropies becomes progressively noisy as a function of coupling strength, size of the subregion A and n. Such a behavior can be understood by assuming an area law, ${S_n \propto \alpha_n(U) \partial A}$, where ${\partial A}$ is the length of the boundary of subsystem A. The quantity we evaluate stochastically is ${\langle Z_n \rangle \propto e^{-(n-1) \alpha_n(U) \partial A}}$. Numerically we see that ${\alpha_n(U)}$ is a growing function of $U$ and weakly dependent on $n$. Hence ${\langle Z_n \rangle}$ decreases exponentially with $U$, $n$ and ${\partial A}$ but the fluctuations do not decrease as quickly such that the signal to noise ratio increases strongly as a function of the above mentioned parameters. As a consequence, and at this point, it is computationally very demanding to pin down the Hubbard $U$ driven transition from the QSH state to the magnetically ordered phase \cite{Hohenadler10,Hohenadler12,Assaad12}. Different sampling schemes, or improved estimators will be required to solve these issues.

We would like to thank L.~Bonnes, P.~Br\"ocker, T.~Grover, I.~Herbut, A.~L\"auchli, R.~Melko and C.~Mudry for discussions. We thank the LRZ-M\"unich and the J\"ulich Supercomputing center for generous allocation of CPU time. Financial support from the DFG grant AS120/9-1 as well as the DFG funded FOR1807 is acknowledged.

\end{document}